\documentclass[12pt]{iopart}

\usepackage{enumerate}
\usepackage{amsthm}

\newcommand{\be}{\begin{equation}}
\newcommand{\ee}{\end{equation}}
\newcommand{\ba}{\begin{eqnarray}}
\newcommand{\ea}{\end{eqnarray}}
\eqnobysec

\begin{document}

\title[An extension of the Newman-Janis algorithm]
{An extension of the Newman-Janis algorithm}

\author{Aidan J Keane}
\address{4 Woodside Place, Glasgow G3 7QF, Scotland, UK.
}

\eads{\mailto{aidan@worldmachine.org}}

\begin{abstract}
The Newman-Janis algorithm is supplemented with a null rotation and applied to the tensors of the Reissner-Nordstr\"om spacetime to generate the metric, Maxwell, Ricci and Weyl tensors for the Kerr-Newman spacetime. This procedure also provides a mechanism whereby the Carter Killing tensor arises from the geodesic angular momentum tensor of the underlying Reissner-Nordstr\"om metric. The conformal Killing tensor in the Kerr-Newman spacetime is generated in a similar fashion. The extended algorithm is also applied to the Killing vectors of the Reissner-Nordstr\"om spacetime with interesting consequences. The Schwarzschild to Kerr transformation is a special case.
\end{abstract}

%\submitto{\CQG}
%\maketitle

\section{Introduction}
In a 1965 paper, Newman and Janis \cite{newman1965} showed that the Kerr metric \cite{kerr63} could be obtained from the Schwarzschild metric using a complex transformation within the framework of the Newman-Penrose formalism \cite{newman1962}. A similar procedure was applied to the Reissner-Nordstr\"om metric to generate the previously unknown Kerr-Newman metric \cite{newman1965b}. For possible physical interpretations of the algorithm see \cite{newman1973,newman1988} and for discussions on more general complex transformations see \cite{flaherty1976,flaherty1980}. More recently, the uniqueness of the Newman-Janis algorithm has been investigated \cite{drake2000}, and exterior calculus applied to facilitate the understanding of Newman-Janis algorithm \cite{ferraro2013}.
It is worth noting that the Kerr-Newman solution can also be derived in a physically straightforward way as the unique solution to the black hole boundary problem \cite{meinel2012}.

The original Newman-Janis algorithm, when applied to the tensors of the Reissner-Nordstr\"om spacetime, only produces the metric tensor for the Kerr-Newman spacetime. The Maxwell, Ricci and Weyl tensors are not given directly by such a simple algorithm: The issue with the Maxwell tensor was acknowledged by Newman \etal \cite{newman1965b}. Furthermore, under the original algorithm the Carter Killing tensor and conformal Killing tensor \cite{carter68,walker70} do not arise as one might expect.

The key to resolving these issues is to arrange for the repeated principal null directions (RPND) of the Reissner-Nordstr\"om spacetime to be mapped to the RPND of the Kerr-Newman spacetime. This is achieved by supplementing the Newman-Janis algorithm with a null rotation. Under this extended Newman-Janis algorithm the Maxwell, Ricci and Weyl tensors for the Kerr-Newman spacetime are obtained from the corresponding objects in the Reissner-Nordstr\"om spacetime.  The Schwarzschild to Kerr transformation is a special case.

It is of interest to investigate the fate of the symmetries under the algorithm. It is known that the Reissner-Nordstr\"om and Kerr-Newman spacetimes are not holomorphically equivalent: This is evident from the fact that the analytic continuation of the Reissner-Nordstr\"om four-dimensional Killing vector algebra is a holomorphic Lie algebra whereas the Kerr-Newman spacetime admits only a two-dimensional algebra \cite{newman1988}. Rather surprisingly, even under the extended algorithm, the timelike and axial Killing vectors in the Reissner-Nordstr\"om spacetime are not mapped directly to those in the Kerr-Newman spacetime. However, as a consequence of the extended algorithm, the two Killing vectors {\it do} arise as members of an algebra generated by the three spherical symmetry Killing vectors of the Reissner-Nordstr\"om spacetime. The extended algorithm also provides a mechanism whereby the Carter Killing tensor arises from the geodesic angular momentum tensor of the underlying Reissner-Nordstr\"om metric. The conformal Killing tensor in the Kerr-Newman spacetime is generated in a similar fashion.

In section \ref{sec:njalgoritm} the extended algorithm is given and compared to the original algorithm.
In section \ref{sec:tensors} the Maxwell, Ricci and Weyl tensors are given explicitly in terms of the derived Newman-Penrose scalars. The Killing vectors, Killing tensor and conformal Killing tensor are considered in section \ref{sec:symmetries}.

\section{An extended Newman-Janis algorithm}\label{sec:njalgoritm}
The Reissner-Nordstr\"om spacetime is given by
\ba
ds^2 & = \xi(r) du^2 + 2 du dr - r^2 (d\theta^2 + \sin^2 \theta d\phi^2)
\label{eq:rn}
\\
\xi(r) & = 1 - r^{-2}(2Mr - e^2).
\label{eq:xi_rn}
\ea
The corresponding metric tensor can be expressed in terms of the {\it complex} null tetrad
\ba
l^a = \delta^a_r, \qquad n^a = \delta^a_u - \case{1}{2} \xi(r) \delta^a_r
\nonumber\\
m^a = {1 \over \sqrt{2}r} (\delta^a_\theta + i \csc \theta \delta^a_\phi)
\label{eq:originalrntetrad}
\ea
as
\be
g^{ab} =  2 l^{(a} n^{b)} - 2 m^{(a} {\bar m}^{b)}.
\label{eq:metrictetrad}
\ee
The extended algorithm is summarized below, consisting of Newman and Janis' algorithm (parts (i) - (iii)) and the null rotation (iv).
The complex substitutions referred to are given in Table \ref{tab:complexsubstitutions}.
\begin{enumerate}
    \item The coordinate $r$ is allowed to take on complex values and the corresponding modified null tetrad $\{ l, n, m, {\bar m} \}$ is constructed. Choosing complex substitutions (b) for $r$ in $m^a$, the tetrad (\ref{eq:originalrntetrad}) is replaced by
    \ba
    l^a = \delta^a_r, \qquad n^a = \delta^a_u - \case{1}{2} \xi^*(r,{\bar r}) \delta^a_r
    \nonumber\\
    m^a = {1 \over \sqrt{2} {\bar r}} (\delta^a_\theta + i \csc \theta \delta^a_\phi)
    \label{eq:tetrad1}
    \ea
    where, choosing complex substitutions (c) and (d) for the $r$ and $r^2$ terms respectively in $\xi(r)$
    \be
    \xi^*(r,{\bar r}) = 1 - (r {\bar r})^{-1}[M (r + {\bar r}) - e^2].
    \label{eq:xi_rn_complexified}
    \ee

    \item Perform the complex coordinate transformation $\{u, r, \theta, \phi \} \mapsto \{ u', r', \theta', \phi' \}$.
        \ba
        u' = u - i a \cos \theta, \qquad r' = r + i a \cos \theta
        \nonumber\\
        \theta' = \theta, \qquad \phi' = \phi
        \label{eq:complexcoordtransf}
        \ea
    on the tetrad $\{ l, n, m, {\bar m} \}$.

    \item Restrict the coordinate $r'$ to be real to obtain the new tetrad $\{ l', n', m', {\bar m}' \}$,
    \ba
    l'^{a} = \delta^{a}_{r'}, \qquad n'^{a} = \delta^{a}_{u'} - \case{1}{2} \xi'(r',\theta') \delta^{a}_{r'}
    \nonumber\\
    m'^{a} = {1 \over \sqrt{2} (r' + ia \cos \theta')} [ i a \sin \theta' (\delta^{a}_{u'} - \delta^{a}_{r'}) + \delta^a_{\theta'} + i \csc \theta' \delta^{a}_{\phi'}]
    \label{eq:tetrad2}
    \ea
    where
    \be
    \xi'(r',\theta') = 1 - [2Mr' - e^2] [(r')^2 + a^2 \cos^2 \theta']^{-1}.
    \ee
    %Under the mapping
    %\be
    %l \mapsto  l', \qquad n \mapsto n', \qquad m \mapsto m', \qquad {\bar m} \mapsto {\bar m}'
    %\label{eq:tetradmapping}
    %\ee
    The new tetrad has the associated metric tensor
    \be
    g'^{ab} =  2 l'^{(a} n'^{b)} - 2 m'^{(a} {\bar m}'^{b)}.
    \label{eq:metrictetrad2}
    \ee

    \item The tetrad vector pair $m'$ and ${\bar m}'$ are not tangent to $r = constant$ surfaces
    but a null rotation (Equation (5.5) of \cite{newman1962}) produces a new pair which are. The null rotation is given by
    \ba
    {\hat l} = l',
    \qquad
    {\hat n} = n' + \alpha {\bar m}' + {\bar \alpha} m' + \alpha {\bar \alpha} l'
    \nonumber\\
    {\hat m} = m' + \alpha l',
    \qquad
    {\hat {\bar m}} = {\bar m}' + {\bar \alpha} l'
    \nonumber\\
    \alpha = i a \sin \theta' (r' + i a \cos \theta')^{-1} / \sqrt{2}.
    \label{eq:complexnullrotation}
    \ea
    This generates a new complex null tetrad $\{ {\hat l}, {\hat n}, {\hat m}, {\hat {\bar m}} \}$
    \ba
    {\hat l}^{a} = \delta^{a}_{r'},
    \qquad
    {\hat n}^{a} = \Sigma^{-1} [(r'^2 + a^2) \delta^{a}_{u'} - \case{1}{2} \Delta \delta^{a}_{r'} + a \delta^{a}_{\phi'}]
    \nonumber\\
    {\hat m}^{a} = {1 \over \sqrt{2} (r' + ia \cos \theta')} [ i a \sin \theta' \delta^{a}_{u'} + \delta^a_{\theta'} + i \csc \theta' \delta^{a}_{\phi'}]
    \label{eq:tetradnullrotation}
    \ea
    where
    \[
    \Sigma = r'^2 + a^2 \cos^2 \theta', \qquad \Delta = r'^2 + a^2 - 2 M r' + e^2.
    \]
    %Under this rotation the tetrads are mapped
    %\be
    %l' \mapsto  {\hat l}, \qquad n' \mapsto {\hat n}, \qquad m' \mapsto {\hat m}, \qquad {\bar m}' \mapsto {\hat {\bar m}}
    %\label{eq:tetradmapping3}
    %\ee
    %and the new tetrad has the associated metric tensor
    The new tetrad has the associated metric tensor
    \be
    {\hat g}^{ab} =  2 {\hat l}^{(a} {\hat n}^{b)} - 2 {\hat m}^{(a} {\hat {\bar m}}^{b)}.
    \label{eq:metrictetrad3}
    \ee
    The metric tensors (\ref{eq:metrictetrad2}) and (\ref{eq:metrictetrad3}) are identical: The null rotation (\ref{eq:complexnullrotation}) is a Lorentz transformation and the metric tensor is invariant under this transformation.
    The metric is the Kerr-Newman metric
    \be
    {\hat g}^{ab} = \Sigma^{-1}
    \left[
    \begin{array}{cccc}
    -a^2 \sin^2 \theta' & (r'^2 + a^2) & 0 & -a \\
    (r'^2 + a^2) & -\Delta & 0 & a \\
    0 & 0 & -1 & 0 \\
    -a & a & 0 & -\csc^2 \theta' \\
    \end{array}
    \right]
    \ee
    The covariant form of the metric is given by
    \be
    \fl {\hat g}_{ab} =
    \left[
    \begin{array}{cccc}
    1 - \Sigma^{-1} (2Mr' - e^2) & 1 & 0 & \Sigma^{-1} (2Mr' - e^2) a \sin^2 \theta' \\
    1 & 0 & 0 & -a \sin^2 \theta' \\
    0 & 0 & -\Sigma & 0 \\
    \Sigma^{-1} (2Mr' - e^2) a \sin^2 \theta' & -a \sin^2 \theta' & 0 & {\hat g}_{\phi \phi} \\
    \end{array}
    \right]
    \ee
    where
    \[
    {\hat g}_{\phi \phi} = -[r'^2 + a^2 - a^2 \sin^2 \theta' \Sigma^{-1} (e^2 - 2Mr') ] \sin^2 \theta'.
    \]
    The complex substitutions for the Maxwell, Ricci and Weyl Newman-Penrose scalars are given in Table \ref{tab:npcomplexsubstitutions}. The corresponding scalars in the Kerr-Newman spacetime, with respect to the tetrad (\ref{eq:tetradnullrotation}), are given by applying the coordinate transformation (\ref{eq:complexcoordtransf}). The associated tensors are obtained from the Newman-Penrose scalars in section \ref{sec:tensors}.

\end{enumerate}
Thus, both the original and extended algorithms provide a satisfactory method for constructing the Kerr-Newman metric tensor.
The notation $l \mapsto  {\hat l}$ shall denote the transformation of $l$ into ${\hat l}$: The first part of this transformation is $l \mapsto  l'$ of the original algorithm (parts (i)-(iii)); the second part of the transformation is the null rotation $l' \mapsto  {\hat l}$.

There are a number of pertinent remarks to be made. First, parts (iii) and (iv) commute in the following sense: If the null rotation with parameter
\be
\alpha = i a \sin \theta' ({\bar r}' + i a \cos \theta')^{-1} / \sqrt{2}.
\label{eq:complexnullrotation2}
\ee
(i.e., (\ref{eq:complexnullrotation}) with $r'$ replaced with ${\bar r}'$)
is carried out prior to the coordinate $r'$ being restricted to be real, then one still obtains the final tetrad (\ref{eq:tetradnullrotation}) and the consequences of the extended algorithm are the same. The significance of this observation lies in the tentative possibility of constructing a complex-dimensional complex manifold, as proposed by Flaherty in Chapter X of \cite{flaherty1976}. However, no further consideration of this will be given here.

Under the extended algorithm the original tetrad is transformed as follows
\be
l \mapsto  {\hat l}, \qquad n \mapsto {\hat n}, \qquad m \mapsto {\hat m}, \qquad {\bar m} \mapsto {\hat {\bar m}}.
\label{eq:extendedalgorithmovralltetradmap}
\ee
In particular, the RPND of Reissner-Nordstr\"om spacetime are transformed into the RPND of the Kerr-Newman spacetime, i.e.,
$l \mapsto  {\hat l}$, $n \mapsto {\hat n}$. This was not the case under the original algorithm. The mapping between the RPND
is not only aesthetically appealing but also crucial to the construction of the Maxwell, Ricci and Weyl tensors in the Kerr-Newman spacetime.
\begin{table}
\caption{\label{tab:complexsubstitutions}Complex substitution types.
The coordinate $r$ appearing in the tensors of the Reissner-Nordstr\"om spacetime can be replaced in one of four ways in part (i) of the algorithm. The four substitution types are listed. The third column gives the resulting real function when the coordinate $r'$ is subsequently taken to be real in part (iii) of the algorithm.}
\footnotesize\rm
\begin{tabular*}{\textwidth}{@{}l*{15}{@{\extracolsep{0pt plus12pt}}l}}
\br
Type & Complex substitution & $r'$ taken to be real\\
\mr

(a) & $r \mapsto r$ & $r = r' - i a \cos \theta \mapsto r' - i a \cos \theta$ \\

(b) & $r \mapsto {\bar r}$ & ${\bar r} = {\bar r}' + i a \cos \theta \mapsto r' + i a \cos \theta$ \\

(c) & $r \mapsto \case{1}{2} (r + {\bar r})$
& $\case{1}{2} (r + {\bar r}) = \case{1}{2} (r' - i a \cos \theta + {\bar r}' + i a \cos \theta) \mapsto r'$ \\

(d) & $r^2 \mapsto r {\bar r}$
& $r {\bar r} = (r' - i a \cos \theta) ({\bar r}' + i a \cos \theta) \mapsto \Sigma$ \\

\br
\end{tabular*}
\end{table}
\begin{table}
\caption{\label{tab:npcomplexsubstitutions}Complex substitutions for the non-zero Newman-Penrose scalars of the Reissner-Nordstr\"om spacetime.}
\footnotesize\rm
\begin{tabular*}{\textwidth}{@{}l*{15}{@{\extracolsep{0pt plus12pt}}l}}
\br
Tensor & Newman-Penrose scalar & Complex substitution type & \\
\mr

Maxwell & $\phi_1 = (e / \sqrt 2) r ^{-2}$ & (a) applied to $r$ term \\

Ricci & $\Phi_{11} = - e^2 r^{-4} / 2$ & (d) applied to $r$ term \\

Weyl & $\Psi_2 = -(M - e^2 r^{-1}) r^{-3}$ & (b) applied to $r^{-1}$ term \\
     & & (a) applied to $r^{-3}$ term \\

\br
\end{tabular*}
\end{table}

\section{Newman-Penrose scalars and tensor fields}
\label{sec:tensors}
The Maxwell, Ricci and Weyl Newman-Penrose scalars, with respect to the tetrad (\ref{eq:tetradnullrotation}), will now be derived explicitly using the extended algorithm. The associated tensors are then obtained from the Newman-Penrose scalars.

The Maxwell scalars for the Reissner-Nordstr\"om spacetime are
\be
\phi_0 = \phi_2 = 0, \qquad \phi_1 = (e / \sqrt 2) r ^{-2}.
\ee
Applying the extended algorithm with the complex substitution (a) of Table \ref{tab:complexsubstitutions} produces the corresponding scalars for the Kerr-Newman spacetime
\be
{\hat \phi}_0 = {\hat \phi}_2 = 0, \qquad {\hat \phi}_1 = (e / \sqrt 2) (r' - ia \cos \theta')^{-2}.
\ee
The Ricci scalars for the Reissner-Nordstr\"om spacetime are
\ba
\Phi_{00} = \Phi_{22} = \Phi_{01} = \Phi_{02} = \Phi_{12} = \Lambda = 0
\nonumber\\
\Phi_{11} = - e^2 r^{-4} / 2.
\ea
Applying the complex substitution (d), the corresponding scalars for the Kerr-Newman spacetime are
\ba
{\hat \Phi}_{00} = {\hat \Phi}_{22} = {\hat \Phi}_{01} = {\hat \Phi}_{02} = {\hat \Phi}_{12} = {\hat \Lambda} = 0
\nonumber\\
{\hat \Phi}_{11} = - e^2 \Sigma^{-2} / 2.
\ea
The Weyl scalars for the Reissner-Nordstr\"om spacetime are
\ba
\Psi_0 = \Psi_1 = \Psi_3 = \Psi_4 = 0
\nonumber\\
\Psi_2 = -(M - e^2 r^{-1}) r^{-3}.
\ea
Applying the appropriate complex substitutions of Table \ref{tab:complexsubstitutions}, the corresponding scalars for the Kerr-Newman spacetime are
\ba
{\hat \Psi}_0 = {\hat \Psi}_1 = {\hat \Psi}_3 = {\hat \Psi}_4 = 0
\nonumber\\
{\hat \Psi}_2 = -[M - e^2 (r' + ia \cos \theta')^{-1}] (r' - ia \cos \theta')^{-3}.
\ea

The corresponding tensors in the Kerr-Newman spacetime can now be easily obtained in terms of the Newman-Penrose scalars and the tetrad basis $\{ {\hat l}, {\hat n}, {\hat m}, {\hat {\bar m}} \}$, see for example Chapter 1 of Chandrasekhar's book \cite{chandrasekhar1983}. The Maxwell tensor is
\ba
{\hat F}^{ab} & = & -{2({\hat \phi}_1 + {\hat {\bar \phi}}_1) {\hat l}^{[a} {\hat n}^{b]}
+ 2({\hat \phi}_1 - {\hat {\bar \phi}}_1) {\hat m}^{[a} {\hat {\bar m}}^{b]}}
\nonumber\\
& = & {2 \sqrt{2} e \Sigma^{-2} [(-r'^2 + a^2 \cos^2 \theta') {\hat l}^{[a} {\hat n}^{b]}}
+ {2 i a r' \cos  \theta' {\hat m}^{[a} {\hat {\bar m}}^{b]}].}
\ea
The Ricci tensor is
\ba
{\hat R}^{ab} & = & - 4 {\hat \Phi}_{11} ({\hat l}^{(a} {\hat n}^{b)} + {\hat m}^{(a} {\hat {\bar m}}^{b)}).
\nonumber\\
& = & 2 e^2 \Sigma^{-2} ({\hat l}^{(a} {\hat n}^{b)} + {\hat m}^{(a} {\hat {\bar m}}^{b)}).
\ea
Since the Ricci scalar ${\hat R} = 0$, the Einstein tensor is also given by the above construction.
Making use of Chandrasekhar's notation \cite{chandrasekhar1983} for vector fields $w$, $x$, $y$, $z$
\ba
\{ w^a x^b y^c z^d \} = & w^a x^b y^c z^d - w^a x^b z^c y^d - x^a w^b y^c z^d + x^a w^b z^c y^d
\nonumber\\
&+ y^a z^b w^c x^d - y^a z^b x^c w^d - z^a y^b w^c x^d + z^a y^b x^c w^d
\ea
the Weyl tensor for the Kerr-Newman spacetime is given by
\ba
{\hat C}^{abcd} = &-({\hat \Psi}_2 + {\hat {\bar \Psi}}_2) (\{ {\hat l}^a {\hat n}^b {\hat l}^c {\hat n}^d \} + \{ {\hat m}^a {\hat {\bar m}}^b {\hat m}^c {\hat {\bar m}}^d \})
\nonumber\\
&+({\hat \Psi}_2 - {\hat {\bar \Psi}}_2) \{ {\hat l}^a {\hat n}^b {\hat m}^c {\hat {\bar m}}^d \}
\nonumber\\
&+ {\hat \Psi}_2 \{ {\hat l}^a {\hat m}^b {\hat n}^c  {\hat {\bar m}}^d \}
+ {\hat {\bar \Psi}}_2 \{ {\hat l}^a {\hat {\bar m}}^b {\hat n}^c  {\hat m}^d \}.
\ea
The explicit expression is rather lengthy and will not be written out in full.

As already acknowledged by Newman \etal \cite{newman1965b}, the Maxwell tensor is not given directly by the original algorithm. This problem occurs because the Maxwell Newman-Penrose scalars behave as follows with respect to each tetrad basis
\ba
\mbox{Reissner-Nordstr\"om} \qquad & \{ l, n, m, {\bar m} \} \qquad & \phi_0 = \phi_2 = 0, \phi_1 \ne 0
\nonumber\\
\mbox{Kerr-Newman} \qquad & \{ l', n', m', {\bar m}' \} \qquad & \phi'_0 = 0, \phi'_2 \ne 0, \phi'_1 \ne 0
\nonumber\\
\mbox{Kerr-Newman} \qquad & \{ {\hat l}, {\hat n}, {\hat m}, {\hat {\bar m}} \} \qquad & {\hat \phi}_0 = {\hat \phi}_2 = 0, {\hat \phi}_1 \ne 0.
\nonumber
\ea
This means that if one is to use the Newman-Janis tetrad $\{ l', n', m', {\bar m}' \}$ one must conjure a {\it non-zero} valued $\phi'_2$ from a $\phi_2 = 0$. However, this problem can be avoided if one is to use the tetrad $\{ {\hat l}, {\hat n}, {\hat m}, {\hat {\bar m}} \}$ as above. Similar arguments apply to the Ricci and Weyl tensors.

\section{Symmetries}\label{sec:symmetries}
An $r$-dimensional isometry algebra will be denoted ${\cal G}_r$.
The Reissner-Nordstr\"om spacetime (\ref{eq:rn}) admits a ${\cal G}_4$ and the Kerr-Newman spacetime a ${\cal G}_2$.
It will be shown that, under the extended algorithm, the timelike and axial Killing vectors in the Reissner-Nordstr\"om spacetime are not transformed directly to those in the Kerr-Newman spacetime. However, under the extended algorithm the two Killing vectors of the Kerr-Newman spacetime do arise as members of an algebra generated by the three spherical symmetry Killing vectors of the Reissner-Nordstr\"om spacetime.

It is well known that the Kerr and Kerr-Newman spacetimes admit both an irreducible Killing tensor and an irreducible conformal Killing tensor \cite{carter68,walker70}. Neither of these arise from the corresponding tensors in the Reissner-Nordstr\"om spacetime via the original algorithm, yet they do under the extended algorithm.

\subsection{Killing vectors}\label{sec:killingvectors}
The Killing vectors of the Reissner-Nordstr\"om spacetime will be expanded in terms of the tetrad $\{ l, n, m, {\bar m} \}$.
Under the extended algorithm this tetrad is then transformed via (\ref{eq:extendedalgorithmovralltetradmap}) to the new tetrad
$\{ {\hat l}, {\hat n}, {\hat m}, {\hat {\bar m}} \}$. The properties of the resulting vector fields are considered.

The Reissner-Nordstr\"om metric admits the ${\cal G}_4$ consisting of the timelike Killing vector
\be
U = \partial_u
\label{eq:rntimelikekv}
\ee
and the $so(3)$ spherical symmetry algebra of Killing vectors with basis
\ba
X_1 = -\cos\phi \partial_\theta + \cot\theta \sin\phi \partial_\phi
\nonumber\\
X_2 = \sin\phi \partial_\theta + \cot\theta \cos\phi \partial_\phi,
\qquad X_3 = \partial_\phi.
\label{eq:rnkv}
\ea
The Lie brackets of the ${\cal G}_4$ are
\ba
[ U,X_1 ] = [ U,X_2 ] = [ U,X_3 ] = 0
\nonumber\\
\left[ X_1,X_2 \right] = X_3, \qquad [ X_2,X_3 ] = X_1, \qquad [ X_3,X_1 ] = X_2.
\ea
The relevant coordinate basis vector fields can be expressed in terms of the complex null tetrad $\{ l, n, m, {\bar m} \}$ of (\ref{eq:tetrad1}) as follows
\ba
\partial_\theta = ({\bar r} m + r {\bar m})/\sqrt 2
\nonumber\\
\partial_\phi = i \sin \theta (r {\bar m} - {\bar r} m)/\sqrt 2
\nonumber\\
\partial_u = n + \case{1}{2} \xi l.
\ea
Thus the Killing vectors of the Reissner-Nordstr\"om spacetime can be written
\ba
U = n + \case{1}{2} \xi l
\nonumber\\
X_1 = {1 \over \sqrt{2}} [-\cos\phi ({\bar r} m + r {\bar m})] + \cot\theta \sin\phi X_3
\nonumber\\
X_2 = {1 \over \sqrt{2}} [\sin\phi ({\bar r} m + r {\bar m})] + \cot\theta \cos\phi X_3
\nonumber\\
X_3 = i \sin \theta (r {\bar m} - {\bar r} m)/\sqrt 2.
\label{eq:rnkv2}
\ea

The vector fields in the Kerr-Newman spacetime corresponding to $\{ U, X_i \}$, $i = 1,2,3$ of the Reissner-Nordstr\"om spacetime can now be determined by employing the extended algorithm. Consider first the vector field
\ba
U = n + \case{1}{2} \xi l.
\ea
The tetrad transformation (\ref{eq:extendedalgorithmovralltetradmap}) and coordinate transformation (\ref{eq:complexcoordtransf})
give rise to the transformations
\be
l \mapsto  {\hat l}, \qquad n \mapsto {\hat n}, \qquad \xi(r) \mapsto \xi'(r', \theta').
\ee
Thus, it is natural to define the vector field associated with $U$ as
\ba
{\hat U} = {\hat n} + \case{1}{2} \xi' {\hat l}.
\ea
Using the identity $\xi'(r', \theta') = \Sigma^{-1} (\Delta - a^2 \sin^2 \theta')$, this vector field can be expressed as
\ba
{\hat U} &= \Sigma^{-1} [(r'^2 + a^2) \partial_{u'} - \case{1}{2} \Delta \partial_{r'} + a \partial_{\phi'}]
+ \case{1}{2} \Sigma^{-1}(\Delta - a^2 \sin^2 \theta') \partial_{r'}
\nonumber\\
&= \Sigma^{-1} [(r'^2 + a^2) \partial_{u'} - \case{1}{2} a^2 \sin^2 \theta' \partial_{r'} + a \partial_{\phi'}].
\ea
Similarly, with the transformation $m \mapsto  {\hat m}$, ${\bar m} \mapsto {\hat {\bar m}}$ and coordinate transformation (\ref{eq:complexcoordtransf}), the vector fields corresponding to the $X_i$ are defined as
\ba
{\hat X}_1 = {1 \over \sqrt{2}} [-\cos\phi' ({\bar r}' {\hat m} + r' {\hat {\bar m}})] + \cot\theta' \sin\phi' {\hat X}_3
\nonumber\\
{\hat X}_2 = {1 \over \sqrt{2}} [\sin\phi' ({\bar r}' {\hat m} + r' {\hat {\bar m}})] + \cot\theta' \cos\phi' {\hat X}_3
\nonumber\\
{\hat X}_3 = i \sin \theta' (r' {\hat {\bar m}} - {\bar r}' {\hat m})/\sqrt 2.
\label{eq:rnkvdefn}
\ea
Explicitly these are
\ba
{\hat X}_1 = -\cos\phi' \partial_{\theta'} + \cot\theta' \sin\phi' {\hat X}_3
\nonumber\\
{\hat X}_2 = \sin\phi' \partial_{\theta'} + \cot\theta' \cos\phi' {\hat X}_3
\nonumber\\
{\hat X}_3 = a \sin^2 \theta' \partial_{u'} + \partial_{\phi'}.
\ea
When the parameter $a = 0$ the vector fields ${\hat U}$, ${\hat X}_i$ reduce to the expressions (\ref{eq:rntimelikekv}) and (\ref{eq:rnkv}) as one would expect. The vector fields ${\hat X}_i$ have Lie brackets
\ba
[ {\hat X}_1,{\hat X}_2 ] = {\hat X}_3 - 2 a {\hat Y}_1, \qquad
[ {\hat X}_2,{\hat X}_3 ] = {\hat X}_1 + 2 a {\hat Y}_2
\nonumber\\
{[} {\hat X}_3,{\hat X}_1 ] = {\hat X}_2 + 2 a {\hat Y}_3
\ea
where
\ba
{\hat Y}_1 = \cos^2 \theta' \partial_{u'},
\qquad
{\hat Y}_2 = \sin \theta' \cos \theta' \sin \phi' \partial_{u'}
\nonumber\\
{\hat Y}_3 = \sin \theta' \cos \theta' \cos \phi' \partial_{u'}.
\ea
The Lie algebra will close with the introduction of three further vector fields
\ba
{\hat Y}_4 = \cos^2 \phi' \sin^2 \theta'\partial_{u'},
\qquad
{\hat Y}_5 = \sin \phi' \cos \phi' \sin^2 \theta' \partial_{u'}
\nonumber\\
{\hat Y}_6 = \partial_{u'}.
\ea
The vector fields ${\hat Y}_I$, $I = 1, \dots,6$ commute amongst themselves. Thus, the vector fields $\{ {\hat X}_i, {\hat Y}_I \}$ are the basis for a nine-dimensional Lie algebra
\ba
[ {\hat X}_1,{\hat X}_2 ] = {\hat X}_3 - 2 a {\hat Y}_1, \qquad
[ {\hat X}_2,{\hat X}_3 ] = {\hat X}_1 + 2 a {\hat Y}_2
\nonumber\\
{[} {\hat X}_3,{\hat X}_1 ] = {\hat X}_2 + 2 a {\hat Y}_3,
\qquad
[ {\hat Y}_I,{\hat Y}_J ] = 0
\nonumber\\
{[} {\hat X}_1,{\hat Y}_1 ] = 2 {\hat Y}_3,
\qquad
{[} {\hat X}_1,{\hat Y}_2 ] = {\hat Y}_5,
\qquad
{[} {\hat X}_1,{\hat Y}_3 ] = - {\hat Y}_1 - {\hat Y}_4
\nonumber\\
{[} {\hat X}_2,{\hat Y}_1 ] = - 2 {\hat Y}_2,
\qquad
{[} {\hat X}_2,{\hat Y}_2 ] = - 2 {\hat Y}_1  + {\hat Y}_4 - {\hat Y}_6,
\qquad
{[} {\hat X}_2,{\hat Y}_3 ] = - {\hat Y}_5
\nonumber\\
{[} {\hat X}_3,{\hat Y}_1 ] = 0,
\qquad
{[} {\hat X}_3,{\hat Y}_2 ] = {\hat Y}_3,
\qquad
{[} {\hat X}_3,{\hat Y}_3 ] = - {\hat Y}_2
\nonumber\\
{[} {\hat X}_1,{\hat Y}_4 ] = - 2 {\hat Y}_3,
\qquad
{[} {\hat X}_1,{\hat Y}_5 ] = - {\hat Y}_2,
\qquad
{[} {\hat X}_1,{\hat Y}_6 ] = 0
\nonumber\\
{[} {\hat X}_2,{\hat Y}_4 ] = 0,
\qquad
{[} {\hat X}_2,{\hat Y}_5 ] = {\hat Y}_3,
\qquad
{[} {\hat X}_2,{\hat Y}_6 ] = 0
\nonumber\\
{[} {\hat X}_3,{\hat Y}_4 ] = -2 {\hat Y}_5,
\qquad
{[} {\hat X}_3,{\hat Y}_5 ] = {\hat Y}_1  + 2 {\hat Y}_4 - {\hat Y}_6,
\qquad
{[} {\hat X}_3,{\hat Y}_6 ] = 0
\label{eq:ninedimalgebra}
\ea
where $I,J = 1, \dots,6$. Note that the combination
\be
a {\hat Y}_1 + {\hat X}_3 = \partial_{u'} + \partial_{\phi'}
\ee
is a Killing vector. The Killing vectors corresponding to the time independence
\be{\hat Y}_6 = \partial_{u'}
\ee
and axial symmetry
\be
a {\hat Y}_1 + {\hat X}_3 - {\hat Y}_6 = \partial_{\phi'}
\ee
in the Kerr-Newman spacetime do not arise directly from those of the Reissner-Nordstr\"om spacetime (i.e., $U$ and $X_3$ respectively), but from the Lie brackets of the nine-dimensional algebra (\ref{eq:ninedimalgebra}) generated by the $so(3)$ spherical symmetry algebra of the Reissner-Nordstr\"om spacetime. It is unknown whether the non-Killing vector fields in the nine-dimensional Lie algebra have any geometrical significance.

The Lie brackets of ${\hat U}$ with the ${\hat X}_i$ does not give a closed algebra.

\subsection{The Killing tensor and conformal Killing tensor}\label{sec:killingtensors}
In the Reissner-Nordstr\"om metric, i.e., in the limit $a = 0$, the Killing tensor reduces to the tensor representing the square of the geodesic angular momentum
\be
J^{ab} = 2 r^2 m^{(a} {\bar m}^{b)}.
\ee
Using (\ref{eq:metrictetrad}) this tensor can expressed as
\be
J^{ab} = 2 r^2 l^{(a} n^{b)} - r^2 g^{ab}.
\ee
Making an appropriate complex substitution for the $r^2$ terms, i.e., (d) and (c) respectively from Table \ref{tab:complexsubstitutions},
the following tensor can be constructed
\be
K^{ab} = 2 r {\bar r} l^{(a} n^{b)} - [\case{1}{2} (r + {\bar r})]^2 g^{ab}.
\label{eq:kt-rn-1}
\ee
As a brief digression, it is noted that by substituting the expression for the metric (\ref{eq:metrictetrad}) into (\ref{eq:kt-rn-1}), this tensor can be written
\ba
K^{ab} & = - \case{1}{2} (r - {\bar r})^2 l^{(a} n^{b)}
- \case{1}{2} (r + {\bar r})^2 m^{(a} {\bar m}^{b)}
\nonumber\\
& = - 2 [im(r)]^2 l^{(a} n^{b)}
- 2 [re(r)]^2 m^{(a} {\bar m}^{b)}
\label{eq:kt-rn-2}
\ea
which displays a complementarity between the two parts.

Under the extended algorithm the Killing tensor in the Kerr-Newman spacetime can be constructed from (\ref{eq:kt-rn-1}): Taking into account the transformation $\{ l, n, m, {\bar m} \} \mapsto \{ {\hat l}, {\hat n}, {\hat m}, {\hat {\bar m}} \}$ and coordinate transformation (\ref{eq:complexcoordtransf}), it is natural to define the Killing tensor to be
\be
{\hat K}^{ab} = 2 \Sigma {\hat l}^{(a} {\hat n}^{b)} - r'^2 {\hat g}^{ab}.
\ee
Explicitly
\[
{\hat K}^{ab} = \Sigma^{-1}
\left[
\begin{array}{cccc}
a^2 r'^2 \sin^2 \theta' & a^2(r'^2 + a^2) \cos^2 \theta' & 0 & a r'^2 \\
a^2(r'^2 + a^2) \cos^2 \theta' & -\Delta a^2 \cos^2 \theta' & 0 & a^3 \cos^2 \theta' \\
0 & 0 & r'^2 & 0 \\
a r'^2 & a^3 \cos^2 \theta' & 0 & r'^2 \csc^2 \theta' \\
\end{array}
\right]
.
\]
The conformal Killing tensor in the Reissner-Nordstr\"om spacetime is
\be
C^{ab} = 2 r^2 l^{(a} n^{b)}.
\label{sec:conformalKillingtensorRN}
\ee
Under the complex substitution (d) in Table \ref{tab:complexsubstitutions}
\be
C^{ab} = 2 r {\bar r} l^{(a} n^{b)}.
\label{sec:conformalKillingtensorRNcomplex}
\ee
The extended algorithm gives the conformal Killing tensor in the Kerr-Newman spacetime, i.e., with the transformation $l \mapsto  {\hat l}$, $n \mapsto {\hat n}$ and coordinate transformation (\ref{eq:complexcoordtransf}) the conformal Killing tensor is defined to be
\be
{\hat C}^{ab} = 2 \Sigma {\hat l}^{(a} {\hat n}^{b)}.
\label{sec:conformalKillingtensorKN}
\ee

\section{Discussion}\label{sec:discussion}
Under the extended algorithm the RPND of the Kerr-Newman spacetime can be constructed from the RPND of Reissner-Nordstr\"om spacetime, i.e.,
\[
l \mapsto  {\hat l}, \qquad n \mapsto {\hat n}.
\]
This, and the consequences outlined in sections \ref{sec:njalgoritm}, \ref{sec:tensors} and \ref{sec:symmetries}, suggest that the null rotation (\ref{eq:complexnullrotation}) is an intrinsic part of the procedure. However, it has to be acknowledged that these results raise as many questions as they answer.

The Schwarzschild to Kerr transformation, i.e., $e = 0$, is a special case and all the above results apply equally well.

The arbitrariness of the complex substitutions present in the original Newman-Janis algorithm remains, and the extended algorithm requires yet more. However, the fact that the tensors {\it can} be constructed in a relatively clean fashion suggests there might be something deeper behind it. It is also interesting that this all relies on the Newman-Penrose formalism: As noted by Flaherty \cite{flaherty1976}, the transformation (\ref{eq:complexcoordtransf}) applied directly to the metric $g_{ab}$ does {\it not} yield the desired results.

It is interesting that, even although the timelike and axial Killing vectors of the Kerr-Newman spacetime do not arise directly from
those in the Reissner-Nordstr\"om spacetime, the extended algorithm still ensures they do arise indirectly as members of a larger algebra generated by the $so(3)$ spherical symmetry algebra of the Reissner-Nordstr\"om spacetime. The significance, if any, of the non-Killing vector fields in the nine-dimensional Lie algebra is unknown.

It is known that the Killing tensor and conformal Killing tensor in the Kerr spacetime {\it do not} arise as the result of any of the standard spacetime symmetries, e.g., conformal Killing vectors, projective, curvature or Weyl collineations, since the Kerr spacetime does not admit any of these (except the two Killing vector fields associated with its axisymmetric
and stationary properties), see \cite{hall2000} for an overview. It remains to determine the geometrical significance of the complex tensor fields (\ref{eq:kt-rn-2}) and (\ref{sec:conformalKillingtensorRNcomplex}).

\section*{Acknowledgments}
I thank Brian Tupper and Ted Newman for useful discussions.

\end{document}